\documentclass{article}
\usepackage{spconf,amsmath,amsfonts, graphicx,multirow}
\usepackage{subcaption}
\usepackage{epstopdf}

\title{A real-time speaker diarization system based on spatial spectrum}
\name{Siqi Zheng, Weilong Huang, Xianliang Wang, Hongbin Suo, Jinwei Feng, Zhijie Yan}
\address{Speech Lab, Alibaba Group\\
\small\texttt{\{zsq174630,yuankai.hwl, xianliang.wxl, gaia.shb, jinwei.feng, zhijie.yzj\}@alibaba-inc.com}}

\begin{document}
\ninept
\maketitle
\begin{abstract}
In this paper we describe a speaker diarization system that enables localization and identification of all speakers present in a conversation or meeting. We propose a novel systematic approach to tackle several long-standing challenges in speaker diarization tasks: (1) to segment and separate overlapping speech from two speakers; (2) to estimate the number of speakers when participants may enter or leave the conversation at any time; (3) to provide accurate speaker identification on short text-independent utterances; (4) to track down speakers movement during the conversation; (5) to detect speaker change incidence real-time. First, a differential directional microphone array-based approach is exploited to capture the target speakers' voice in far-field adverse environment. Second, an online speaker-location joint clustering approach is proposed to keep track of speaker location. Third, an instant speaker number detector is developed to trigger the mechanism that separates overlapped speech. The results suggest that our system effectively incorporates spatial information and achieves significant gains. 

\end{abstract}
\section{Introduction}
\label{sec:intro}

Speaker diarization is a process of finding the optimal segmentation based on speaker identity and determining the identity of each segments. It is also commonly stated as a task to answer the question ``who speaks when". It tries to match a list of audio segments to a list of different speakers. Unlike the speaker verification task, which is a simple one-to-one matching, speaker diarization matches $M$ utterances to $N$ speakers, and, in some situations, $N$ is often unknown. 

Agglomerative hierarchical clustering (AHC) over speaker embeddings extracted from neural networks has become the commonly accepted approach for speaker diarization\cite{DBLP:conf/icassp/WangDWMM18}\cite{DBLP:conf/icassp/Garcia-RomeroSS17}\cite{DBLP:conf/icassp/ZhangWZP019}. Variational Bayes (VB) Hidden Markov Model has proven to be effective in many studies\cite{DBLP:journals/taslp/DiezBLC20}\cite{DBLP:conf/icassp/LandiniWDBMZMSP20}\cite{DBLP:conf/odyssey/DiezBM18}. A refinement process is usually applied after clustering. VB resegmentation is often selected as means for refinement\cite{DBLP:conf/icassp/LandiniWDBMZMSP20}\cite{DBLP:conf/icassp/SellG15}.

The real-time requirement poses another challenge for speaker diarization \cite{DBLP:conf/icassp/AronowitzZ20}. To be specific, at any particular moment, it is required that we determine whether a speaker change incidence occurs at the current frame within a delay of less than 500 milliseconds. This restriction makes refinement process such as VB resegmentation extremely difficult. Since we can only look 500 milliseconds ahead, the speaker embedding extracted from the speech within that short window can be biased. Therefore, finding the precise timestamp of speaker change in a real-time manner still remains quite intractable for conventional speaker diarization technique. In this work we try to seek solutions from spatial information. As it is shown in later sections, the additional spatial information has been proven to be effective. 

DER is a widely-used metric for measuring the performance of speaker diarization systems\cite{DBLP:conf/interspeech/Galibert13a}. In this work we will continue to use it as one of the measurement metrics. In addition, we introduce another metric to emphasize the importance of finding the accurate point in time of a speaker turn. When attaching to an Automatic Speech Recognition system, it is crucial that the estimated speaker change timestamp is precisely located within the interim between two speaker utterances. Missing the interim by a few frames may not result in a noticeable degradation in DER, but it may cause the mis-classification of the first or last words in a sentence. This can be harmful to the ASR system as well as the final user experience. Hence we introduce segmentation accuracy, which measures the percentage of estimated speaker change timestamp that correctly lies inside the interim between two speaker utterances.

Noise in far-field environment has been a long-standing challenge for speaker verification and diarization. Microphone array provides effective solutions for far-field sound capture and sound source localization in adverse environment. Beamforming-based solutions work as a spatial filtering to enhance the signal from desired direction and suppress the signal from undesired direction \cite{van1988beamforming}.

Microphone array contributes to our speaker diarization system in two ways. First, the ability to localize sound source enables the system to find the optimal segmentation points with remarkable accuracy. Locations of each segments are effective complements for speaker embeddings in joint clustering, especially for short segments. Second, differential directional microphone array significantly improves the quality of speaker's voice in far-field, noisy environment, which in turn enhances the representative power of speaker embeddings. 

\section{System Description}
\begin{figure*}[t]
  \centering
  \includegraphics[width=\linewidth]{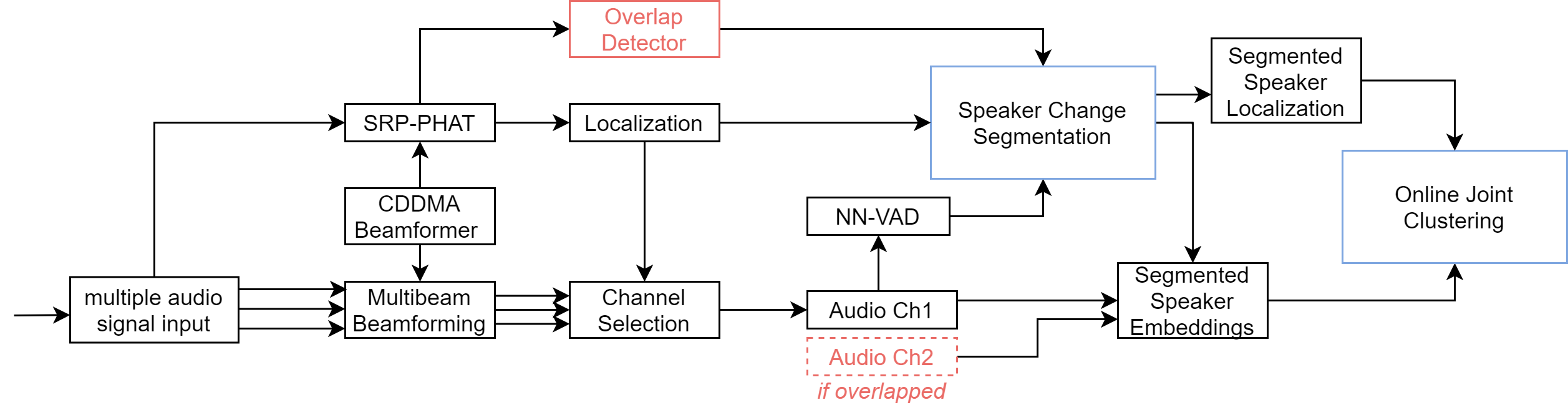}
  \caption{The microphone array signals are exploited to generate spatial information for localization and multiple spatially separated signal based on CDDMA beamformers. The optimal channels are selected based on localization results. An NN-VAD is estimated from the selected channel. The results of NN-VAD, along with localization of each frame, determine the occurrence of speaker change and hence enable segmentation. Speaker embeddings are then extracted from segmented audios. Finally, a joint online clustering is performed to cluster speaker embedding and their corresponding localization.}
  \label{fig:system}
\end{figure*}

In this section we describe our proposed speaker diarization system with spatial spectrum (SDSS). First we describe our design of microphone array and discuss how it helps to capture speaker's voice in far-field noisy environment. Microphone array also provides information to estimate the localization, or direction-of-arrival (DOA), of the sound source. The speaker segmentation is jointly determined by the source localization and a neural network-based voice activity detection (NN-VAD). Then the discussion moves on to our proposed online joint clustering approach based on collaborative contributions of speaker embeddings and localization. A phonetically-aware coupled network that minimizes the impacts of mismatch of speech content is investigated. Finally, we propose an overlap detector that helps to detect and separate overlapping speech. Figure \ref{fig:system} provides an illustration of the system.

\subsection{Audio segmentation based on beamforming}

The audio segmentation and finding the exact point in time of a speaker change incidence are determined by the joint efforts of spatial localization and NN-VAD.

Like a multi-beam system in the previous work \cite{chen2017cracking}, a set of differential beamformers are designed to attract the signal from different directions. More specifically, the look-direction of each beamformer is uniformly distributed around a circle to cover the whole space. The output signals of the beamformers are spatially separated from each other. 

\begin{figure}[h!] 
\centering
  \includegraphics[width=0.8\linewidth]{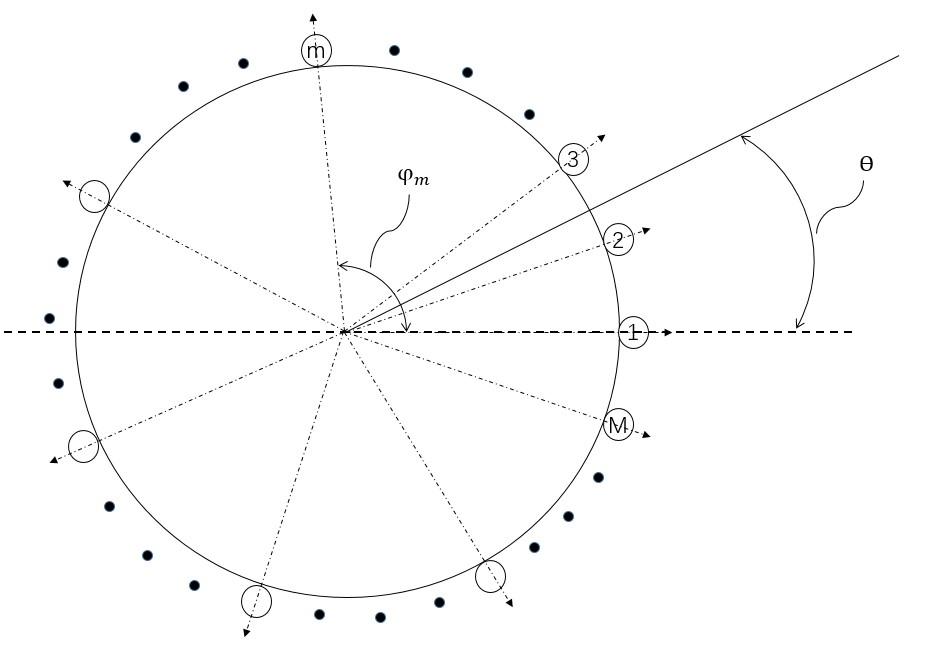}
  \caption{Uniform circular array with directional microphones}
  \label{micarray}
\end{figure}

In order to achieve better performance, we utilize the previously proposed circular differential directional microphone array (CDDMA) design \cite{huangdifferential}, where the microphone array is based on a uniform circular array with directional microphones depicted in Figure \ref{micarray}. All the directional elements are uniformly distributed on a circle and directions are pointing outward. The CDDMA beamformer is given as below:
\begin{equation}\label{solution}
	\textbf{h}_{cddma}(\omega)  = \textbf{R}^{H}( \omega, \boldsymbol{\theta }) [\textbf{R}( \omega, \boldsymbol{\theta }) \textbf{R}^{H}( \omega, \boldsymbol{\theta })]^{-1} \textbf{c}_{\boldsymbol{\theta }} . 
\end{equation}
where the vector $\textbf{c}_{\boldsymbol{\theta }}$ defines the acoustic properties of the beamformer such as beampattern; the constraint matrix $\textbf{R}( \omega, \boldsymbol{\theta })$ of size $\mathit{N} \times \mathit{M} $ is constructed by the directional microphone steering vector which exploits the acoustics of microphone elements. As proven in \cite{huangdifferential}, the CDDMA-beamformer demonstrates significant advantages over the conventional method in terms of white noise gain (WNG) and directivity factor (DF), two commonly used performance measures for differential beamforming. WNG measures the efficacy to suppress spatially uncorrelated noise for beamformers \cite{brandstein2013microphone}. And DF quantifies how the microphone array performs in the environment of reverberation\cite{benesty2018fundamentals}.

For the sound source localization task, we utilize the SRP-PHAT algorithm \cite{dibiase2000high} with the CDDMA-beamformer described above. At first, we assume that the microphone array signals at $n^{th}$ frame are received as:
\begin{equation}\label{eq:1}
\textbf{x}\left( \omega, \theta \right) = [x_{1}, x_{2},\ \cdots \ x_{M}]^{T},
\end{equation}
where the superscript $^{T}$ represents the transpose operator, $\omega = 2 \pi \mathit{f} $ is the angular frequency, $\mathit{f}$ is the temporal frequency,  $\theta$ is the incident angle of the signal, and $x_{m}$ represents the signal of each microphone. Then, for each candidate of incident angle $\theta$, we design each corresponding CDDMA-beamformer to target at the direction of $\theta$, denoted as $\textbf{h}_{cddma}(\omega, \theta)$, and we calculate the transient steering response power(SRP) at $n^{th}$ frame as below:
\begin{equation}\label{eq:1}
\textbf{P}_{n} \left(  \theta \right) = \int_{-\infty}^{+\infty} | \textbf{x}\left( \omega, \theta \right)^{H} \textbf{h}_{cddma}(\omega, \theta)|^{2} d \omega.
\end{equation}
In practice, we will take a recursive smoothing of SRP to obtain the estimate as below:
\begin{equation}\label{eq:1}
 \hat{ \textbf{P} }_{n} \left(  \theta \right) = \alpha  
 \hat{ \textbf{P} }_{n-1} \left(  \theta \right) + (1 - \alpha) \textbf{P}_{n} \left(  \theta \right) 
\end{equation}
where $\alpha$ is called the forgetting factor, ranging from zero to one. The locale estimate of incident angle for current frame is given by:
\begin{equation}\label{eq:1}
\hat{\theta} = \underset{\theta}{argmax} \ \textbf{P}\left(  \theta \right)
\end{equation}
Based on the estimate of SRP at each frame, we can form a spatial spectrum  as below:
\begin{equation}\label{eq:2}
\mathcal{B} \left(  \theta , n \right) = \hat{\textbf{P}}_{n} \left(  \theta \right), \forall \ n \in \mathbb{N}^+
\end{equation}

A smoothing function is applied to the estimated angle $\hat\theta$ of each frame to filter noisy angles. In this case we apply a medium filter to the small window centered by the current frame. The filtered angle goes through an online clustering one after another. Every time an angle incidence that lies outside of the current cluster is spotted, we marks that current frame as a possible speaker change timestamp.

\subsection{Speaker diarization}

Consider two utterances $u_A$ and $u_B$. Traditional speaker verification task tries to estimate the probability that $u_A$ and $u_B$ are from the same speaker,
$$P(same|u_A,u_B).$$
With additional estimated direction-of-arrival(DOA) $d_A$ and $d_B$, assuming that they are independent from $u_A$ and $u_B$, we instead try to estimate the joint conditional probability
$$P(same|u_A,u_B)P(same|d_A,d_B).$$

It is worthwhile to mention that even though the utterances $u_A$ and $u_B$ are generated by the beam related to DOA $d_A$ and $d_B$, their effects are marginal from the perspective of speaker verification. Hence we find that it is safe to make the independence assumption for the sake of this task.

$P(same|d_A,d_B)$ is obtained from DOA identity estimator learned from the train set. The parameters of DOA identity estimator are estimated by a linear classification model, in polar coordinates. 

An online agglomerative hierarchical clustering is performed on the audio segments and source location, based on the joint conditional probability. While the traditional clustering updates only the speaker embedding centroid, we also keep updating the localization corresponding to that centroid. This allows us to track down speaker movement during the conversation. We introduce a hyperparameter $T_d$ to identify the occurrence of such movement.

To formalize, let $S = \{s_1,s_2,\ldots, s_N\}$ be the list of current active speakers in the session and $D = \{d_1,d_2,\ldots,d_N\}$ be the corresponding location of these speakers. Let $u'$  be a new incoming speech segment and and $d'$ be the source localization of that speech segment. We define the following probability:

$p_{new} = P(u' \not\in s, \forall s\in S| S,D,u',d')$,

$p_{update} = P(u' \in s, s \in S, \angle(d_s,d')< T_d | S,D,u',d')$,

$p_{move} = P(u' \in s, s \in S, \angle(d_s,d') \geq T_d| S,D,u',d')$.

When running real-time, we update the state of conversation $(S,D)$ every time we receive a new sample segment $(u',d')$. Other than the three actions listed above, another option the system may take is to estimate a most likely speaker identification without making any updates to the state $(S,D)$, when the confidence is not high enough.

\subsection{Phonetically-aware coupled network}

The CDDMA design significantly improves the quality of the capture of speaker's voice in far-field environment. Noise from other directions are also effectively suppressed. This greatly ameliorates two major sources of negative impacts in speaker diarization or verification. There is one other factor that affects the performance of speaker diarization - the mismatch of content of speech among segments.

A major challenge for the speaker clustering is that the embeddings are extracted from length-varied segments of random speech, some of which are relatively short. Speaker verification on short text-independent utterances is difficult because the content of speech introduces a fair amount of bias to the estimated speaker embedding. 

Channel mismatch is less of a concern in our task not only because microphone array has significantly reduced noise, but also because in diarization all segments are collected in the same environment within the same time session. Therefore we place more focus on dealing with the impacts of phonetic contents of the speech, especially for short utterances.

Phonetically-aware coupled network(PacNet) as described in \cite{DBLP:conf/interspeech/ZhengS20} is implemented. PacNet focuses on normalizing the effects of phonetic contents by directly comparing both the phonetic and acoustic information of the two utterances. The contrastive training scheme ensures that the effects of phonetic contents are minimized.

\subsection{Separation of overlapping speech}

Beamforming technique discussed above allows us to separate signals from different DOA. Another problem that needs to be addressed is to determine the number of valid speech sources at the given moment.

An HMM-based state estimator is used to determine the number of speech sources in a streamlined manner. The estimator consists of three output states: 2 speech sources, 1 speech source, and no speech. The transition probabilities between states are jointly determined by the frame posterior of NN-VAD and the result of unimodality test. While it is straightforward to see how frame posterior of NN-VAD assists us in determining the existence of valid speech, the need for a unimodality test may require more explanation. For any particular frame, if we map the SRP-PHAT value against the corresponding direction we obtain a curve peaked at the sound source. At time frame $t$, the curve is generated by the function $\mathcal{B} \left(  \theta , t \right)$. Clean, single speech source signals often result in a bell-shaped unimodal curve. A unimodality test statistics is an effective indicator for the detection of overlapping speech. We use Hartigan's Dip Test to measure the unimodality \cite{hartigan1985dip}. To be specific, we measure the maximum difference over all DOAs, between the empirical distribution of SRP-PHAT, and the unimodal distribution function that minimizes that maximum difference. If there are more than 2 valid speech sources, we only localize the two most significant sources.   

\section{Experiments}

\begin{table*}[tb]
    \caption{Accuracy of finding the precise timestamp of speaker change in various situations, where the interim is 100ms and 300ms, respectively. Evaluation is performed on Part II of the evaluation set.}
    \centering
    
    \begin{tabular}{lcccccc}
    \hline
      \multirow{2}{*}{Background noise}  & \multicolumn{2}{c}{xvector real-time, no spatial information} &  \multicolumn{2}{c}{xvector offline, no spatial information} & \multicolumn{2}{c}{our system - SDSS} \\
       & \small{100ms}& \small{300ms} & \small{100ms} & \small{300ms}  &\small{100ms}& \small{300ms} \\
      \hline
       White noise  & 46.26\% & 57.34\% & 77.34\% & 85.1\%  & 100.0\% & 100.0\%\\ \hline
    Fan Blowing & 40.33\%  & 51.97\% & 61.2\% & 69.38\%  & 99.17\% & 100.0\%\\ \hline
    Knock on the table & 32.18\% & 50.85\% & 49.8\% & 56.62\%  & 96.7\% & 99.3\%\\ \hline
    Keyboard Noise & 35.59\% & 55.0\% & 60.46\% & 64.53\%   & 98.26\% & 100.0\% \\ \hline
    Chair Sliding& 38.54\%& 53.67\% & 62.18\% & 67.0\% & 97.9\%  & 98.5\% \\ \hline
    \end{tabular}
    \label{tab:Result1}
\end{table*}

\subsection{Corpus}

The evaluation set we use consists of audios recorded and processed by the microphone array design CDDMA we mentioned in the previous section. The data set includes two parts. Part I is collected from the actual meeting conversations. Speaker segments and identifications are manually labelled. We recorded 30 speakers from 10 meetings. Each of the meeting lasts 8 to 30 minutes and has somewhere between 3 to 6 participants. The duration of the speaker-segments range from 1 second to several minutes, with the majority to be under 10 seconds.  While most of the participants stay in one position during the meeting, about 5 percent of the speakers would stand up and move to other seats at some point during the session. 

Part II of evaluation set consists of simulated data recorded in the studio. This part of the data set is used to measure the segmentation accuracy - the proportion the estimated speaker change timestamps rest precisely within the interim between two speakers. Hence, the duration and timestamp of the interim are strictly controlled and documented as ground truth reference. Audios of interim durations of 100ms and 300ms are simulated, respectively. Three high resolution speakers are placed around the microphone array, forming an angle of 120 degree with one another. During simulation, each of the speakers play recorded voices of human in turn, with the fixed length and timestamps of the interims. In addition, we simulate noise that are commonly seen in the actual meeting rooms, such as the noise of blowing fans, keyboards, chair sliding, and knocks no the tables. 

For train set we use VoxCeleb dataset \cite{DBLP:conf/interspeech/NagraniCZ17}, since we have not collected enough labelled data from our microphone-array design to train a speaker embedding network.

\subsection{Experimental Setup}

Three experiments are conducted to verify the performance of our system. First, we compare the CDDMA beamformer in our system to two widely-used fixed beamformers: conventional circular differential microphone array (CDMA) \cite{benesty2015design} and Delay-and-Sum beamformer applied in BeamformerIt \cite{anguera2007acoustic}. The simulation setup is using 12 elements to form an array with diameter of 8 cm, where the first-order cardioid are used for CDDMA and omni-directional microphones are used for others. And the CDDMA and CDMA beamformer is designed to target at a super-cardioid beampattern with parameters
of $c_{\theta=3/4\pi} = c_{\theta=5/4\pi} = 0$, and $c_{0}=1$ in equation (\ref{solution}). 

In the second experiment we compare the accuracy of the estimated timestamps of speaker change. The baseline system is the conventional x-vector-based speaker diarization. The detailed configurations are set up as described in the Kaldi recipe\cite{Povey:192584}. The experiment is conducted on Part II of the evaluation set.

The third experiment, run on part I of the evaluation set, compares the DER between several systems on real meeting conversations. 

\section{Results \& Discussions}
\begin{figure}
\centering

\begin{subfigure}{.5\linewidth}
  \centering
  \includegraphics[width=\linewidth]{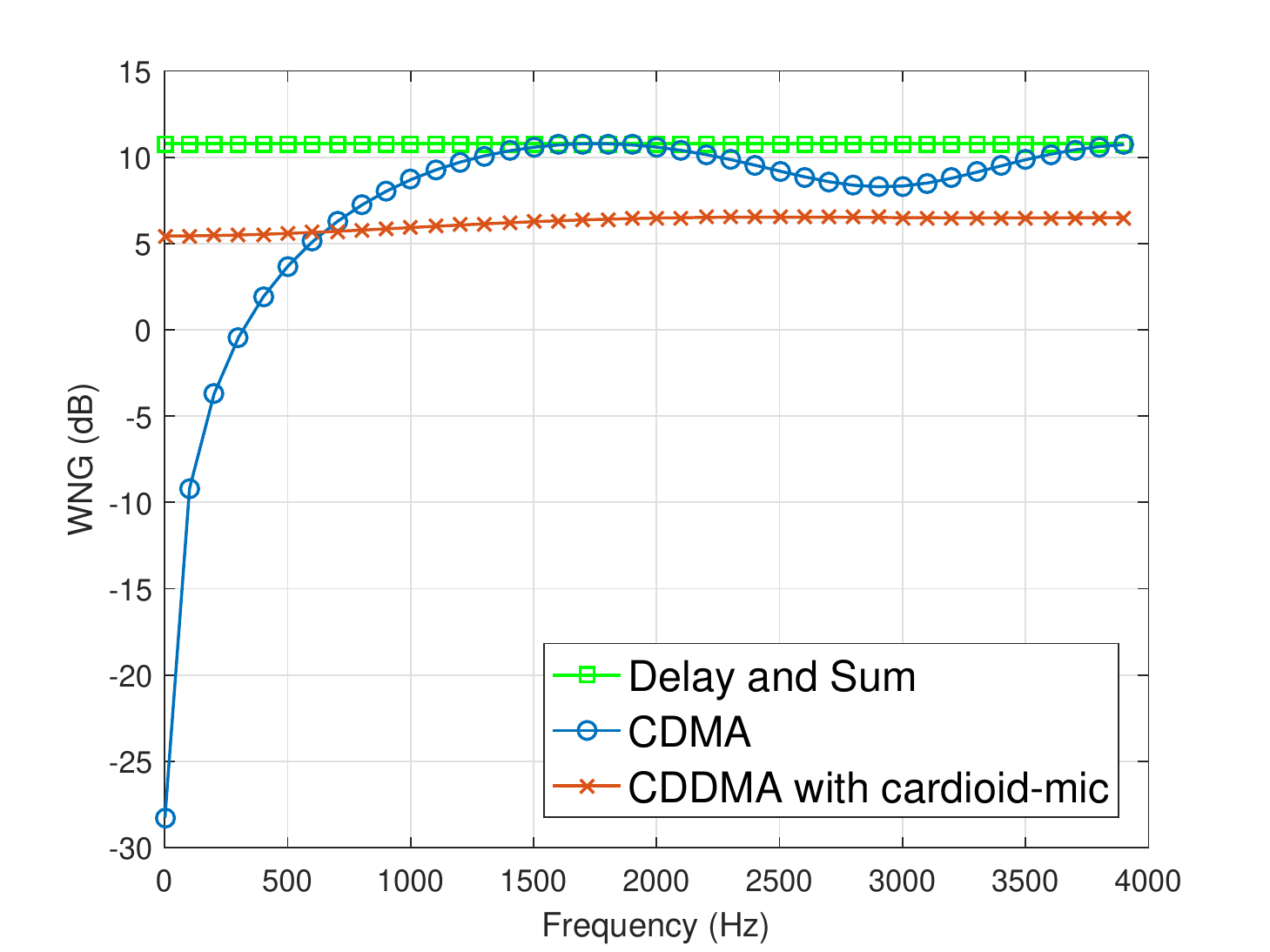}
  \caption{WNG}
  \label{fig:sub1}
\end{subfigure}%
\begin{subfigure}{.5\linewidth}
  \centering
  \includegraphics[width=\linewidth]{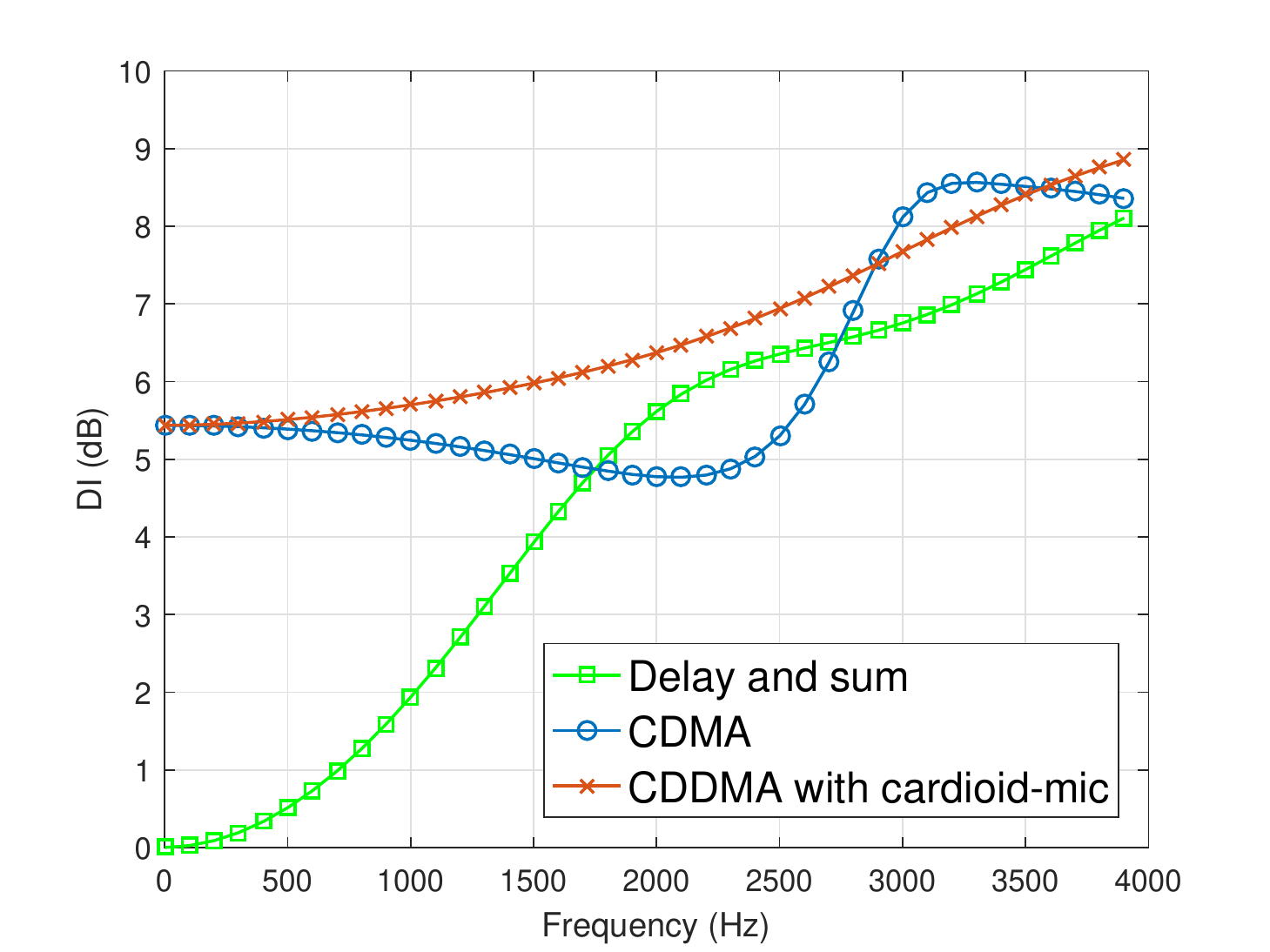}
  \caption{DI}
  \label{fig:sub2}
\end{subfigure}
\caption{WNG and DI comparison among Delay-and-Sum, CDMA and CDDMA
}
\label{fig:test}
\end{figure}
Figure \ref{fig:test} displays the WNG and DI comparison results between CDDMA, CDMA and Delay-and-Sum beamformer. The conventional CDMA suffers from the significant decrease of WNG at low frequency. This causes  white noise amplification in practice. Both the Delay-and-Sum beamformer and CDDMA outperform CDMA in terms of WNG. Although the Delay-and-Sum beamformer can achieve the highest WNG, it has very low DI at low frequency. This reduces the spatial filtering ability of Delay-and-Sum beamformer on overlapping and noisy speech. Therefore, we utilize the recently-proposed CDDMA beamformer in our system to achieve a balanced performance for our tasks.

An interim is the duration between the speech ending frame of previous speaker and the starting point of next speaker. Table \ref{tab:Result1} measures the segmentation accuracy of various systems. An estimated change point that misses the interim by several frames may result in the misclassification of the first or last few words of a sentence. As we can see, without spatial information, the accuracy of finding the correct timestamp of speaker change is relatively low, ranging from $30\%$ to $70\%$, depending on the environment. With the aid of spatial localization, we are able to achieve a decent accuracy close to $100\%$.

We also observed significant improvement in DER. As shown in Table \ref{tab:DER}, SDSS-x-vector system reduces the DER from $23.17\%$ to $6.11\%$, which demonstrates the power of spatial information on speaker diarization. Normalizing speech contents using PacNet further reduces the DER to $5.22\%$.

\begin{table}[h]
    \caption{DER comparison of various systems, evaluated on Part I of the evaluation set.}
    \centering
    \begin{tabular}{|c|c|c|}
    \hline
      System   & with spatial information & DER(\%) \\
      \hline
    x-vector online & No & 23.17 \\ \hline
    SDSS-x-vector & Yes & 6.11 \\ \hline
    SDSS-PacNet & Yes &\textbf{5.22} \\
    \hline
    \end{tabular}
    \label{tab:DER}
\end{table}

\section{Conclusion}

In this paper we propose a speaker diarization system that effectively incorporates spatial information and achieves remarkable performance gains compared to the traditional speaker diarization system without the assistance of microphone array.

In this study we aim provide an experimental knowledge of integrating microphone array technology in speaker diarization system and find out where the horizon is at. The convincing results suggest a promising path for large-scale real-time application of speaker diarization.

As the costs of microphone array continuously decrease, we witness the increase in popularity of this technology. Considering the remarkable gains and benefits spatial information brings to speaker diarization system, it is worthy to incorporate microphone array in certain scenarios.

\bibliographystyle{IEEEbib}
\bibliography{./strings,./refs}

\end{document}